\documentclass[reprint,aps,prb,twocolumn,superscriptaddress]{revtex4-2}

\usepackage{graphicx}
\usepackage{amsfonts}
\usepackage{amsmath}
\usepackage{amssymb}
\usepackage{bbm}
\usepackage{blindtext}  
\usepackage{bm}
\usepackage{braket}
\usepackage{bbold}
\usepackage{color}
\usepackage{comment}
\usepackage{hyperref}
\usepackage{latexsym}
\usepackage{multirow}
\usepackage{subfigure}
\usepackage{soul}
\usepackage{tikz}
\usepackage{tabularx}
\usepackage{upgreek}
\usepackage{verbatim}
\usepackage{xcolor}
\usepackage{physics}

\hypersetup{
    colorlinks=true,
    linkcolor=blue,
    urlcolor=blue,
    filecolor=blue,
    citecolor=blue
    }
\usepackage{siunitx}

\begin{document}

\title{Realizing the Emery Model in Optical Lattices for Quantum Simulation\\ of Cuprates and Nickelates}

\author{Hannah Lange}
\thanks{These authors contributed equally to this work}
\affiliation{Department of Physics and Arnold Sommerfeld Center for Theoretical Physics (ASC), Ludwig-Maximilians-Universit\"at M\"unchen, Theresienstr. 37, M\"unchen D-80333, Germany}
\affiliation{Munich Center for Quantum Science and Technology (MCQST), Schellingstr. 4, M\"unchen D-80799, Germany}

\author{Liyang Qiu}
\thanks{These authors contributed equally to this work}
\email{liyang.qiu@mpq.mpg.de}
\affiliation{Max-Planck-Institute for Quantum Optics, Hans-Kopfermann-Str.1, Garching D-85748, Germany}
\affiliation{Munich Center for Quantum Science and Technology (MCQST), Schellingstr. 4, M\"unchen D-80799, Germany}

\author{Robin Groth}
\affiliation{Max-Planck-Institute for Quantum Optics, Hans-Kopfermann-Str.1, Garching D-85748, Germany}
\affiliation{Munich Center for Quantum Science and Technology (MCQST), Schellingstr. 4, M\"unchen D-80799, Germany}

\author{Andreas von Haaren}
\affiliation{Max-Planck-Institute for Quantum Optics, Hans-Kopfermann-Str.1, Garching D-85748, Germany}
\affiliation{Munich Center for Quantum Science and Technology (MCQST), Schellingstr. 4, M\"unchen D-80799, Germany}

\author{Luca Muscarella}
\affiliation{Max-Planck-Institute for Quantum Optics, Hans-Kopfermann-Str.1, Garching D-85748, Germany}
\affiliation{Munich Center for Quantum Science and Technology (MCQST), Schellingstr. 4, M\"unchen D-80799, Germany}

\author{Titus~Franz}
\affiliation{Max-Planck-Institute for Quantum Optics, Hans-Kopfermann-Str.1, Garching D-85748, Germany}
\affiliation{Munich Center for Quantum Science and Technology (MCQST), Schellingstr. 4, M\"unchen D-80799, Germany}

\author{Immanuel Bloch}
\affiliation{Max-Planck-Institute for Quantum Optics, Hans-Kopfermann-Str.1, Garching D-85748, Germany}
\affiliation{Munich Center for Quantum Science and Technology (MCQST), Schellingstr. 4, M\"unchen D-80799, Germany}
\affiliation{Department of Physics, Ludwig-Maximilians-Universit\"{a}t, 80799 Munich, Germany}

\author{Fabian Grusdt}
\affiliation{Department of Physics and Arnold Sommerfeld Center for Theoretical Physics (ASC), Ludwig-Maximilians-Universit\"at M\"unchen, Theresienstr. 37, M\"unchen D-80333, Germany}
\affiliation{Munich Center for Quantum Science and Technology (MCQST), Schellingstr. 4, M\"unchen D-80799, Germany}

\author{Philipp M. Preiss}
\affiliation{Max-Planck-Institute for Quantum Optics, Hans-Kopfermann-Str.1, Garching D-85748, Germany}
\affiliation{Munich Center for Quantum Science and Technology (MCQST), Schellingstr. 4, M\"unchen D-80799, Germany}

\author{Annabelle Bohrdt}
\affiliation{Department of Physics and Arnold Sommerfeld Center for Theoretical Physics (ASC), Ludwig-Maximilians-Universit\"at M\"unchen, Theresienstr. 37, M\"unchen D-80333, Germany}
\affiliation{Munich Center for Quantum Science and Technology (MCQST), Schellingstr. 4, M\"unchen D-80799, Germany}

\date{\today}
\begin{abstract}
The microscopic origin of high-temperature superconductivity in cuprates remains one of the central open questions in condensed matter physics. Growing experimental and theoretical evidence suggests that the bare single-band Fermi-Hubbard model may not fully capture properties of cuprates such as superconductivity, motivating us to revisit the canonical three-band model of the copper-oxide planes -- the Emery model -- from which the single-band counterpart was originally derived. Here, we propose and analyze a quantum simulation scheme for realizing the Emery model in regimes relevant to cuprates and infinite-layer nickelates with today's ultracold atom quantum simulation platforms, enabling the exploration of the three-band physics on system sizes that are challenging for current numerical methods. Specifically, we show that a two-dimensional optical lattice with a superimposed pattern of repulsive potentials can be designed to study low-temperature properties for variable parameter regimes of the Emery model relevant to cuprates as well as infinite-layer nickelates. Our results pave the way for real material simulations with ultracold atom quantum simulators and a better understanding of the physics of unconventional superconductors.
\end{abstract}

\maketitle

\begin{figure}[t]
    \centering
    \includegraphics[width=0.49\textwidth]{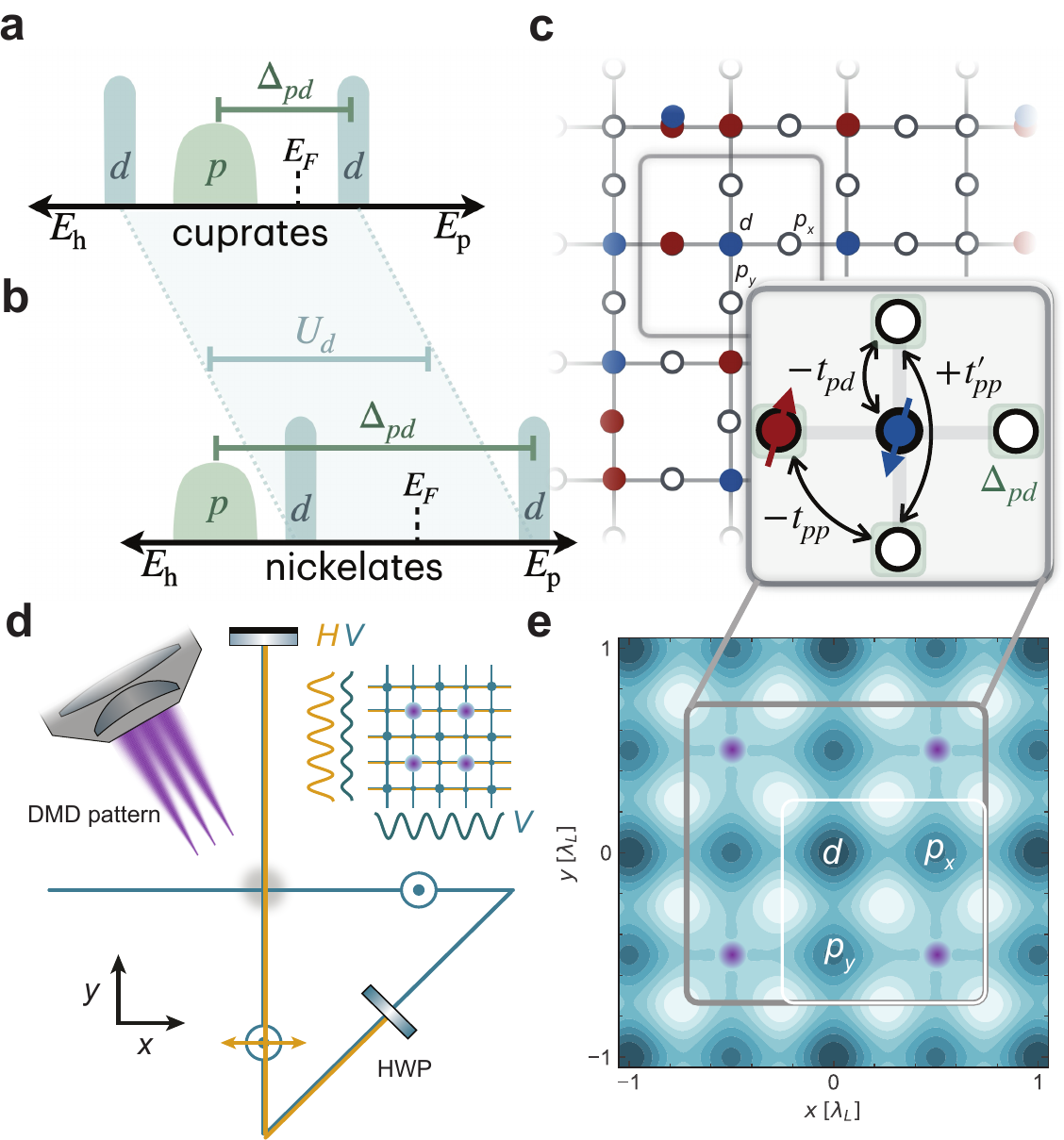}
    \caption{The three-band Emery model of cuprates in cold atom quantum simulators: Band structure of cuprates (\textbf{a}) and nickelates  (\textbf{b}) with copper $d$ and oxygen $p$ bands. Performing a particle-hole transformation effectively corresponds to reversing the sign of the energy $E_\mathrm{p}\to E_\mathrm{h}$. \textbf{c)} The $d$ and $p_x, p_y$ orbitals also appear in a lattice description of the copper-oxide layers in cuprates, with hopping strengths $t_{pd}$, $t_{pp}^{(\prime)}$. In a description for holes, $p_x, p_y$ orbitals are offset by $\Delta_{pd}>0$. \textbf{d)}
    Experimental setup: A single laser beam traverses the atomic cloud (gray) sequentially along the $x$ and $y$ axes and is retro-reflected to form a 2D lattice. By inserting a half waveplate (HWP) between the first and second passes, the polarization of the lattice along the $y$-axis can be dynamically tuned from the initial input polarization ($V$, green) to a fully orthogonal state ($H$, orange). As illustrated in the top-right inset, the resulting passively phase-stable lattice features a programmable offset between nearest-neighbor sites. By using a Digital Micromirror Device (DMD) to project repulsive local potentials (violet arrays) through the objective -- blocking out every fourth site -- a dynamically controllable potential for the Emery model is generated (\textbf{e}). 
    }
    \label{fig:1}
\end{figure}

 Since the discovery of cuprate-based superconductors over four decades ago~\cite{Bednorz1986}, unraveling the microscopic mechanisms underlying high-temperature superconductivity remains one of the most prominent unresolved problems in condensed matter physics. Extensive research has sought to understand the underlying electronic properties of these strongly correlated materials. The electronic degrees of freedom in the copper-oxide layers can be modeled by a three-band description on a Lieb lattice geometry, the Emery model \cite{Emery1987}, involving a copper and two oxygen sites per unit cell with an energy gap $\Delta_{pd}$, see Fig.~\ref{fig:1}. By constructing the Wannier functions centered at the copper-sites, the Emery model can be downfolded to a simplified description in terms of the famous single-band Fermi-Hubbard model \cite{Zhang1988}, which has been studied extensively in the last decades \cite{Arovas_2022}. However, increasing evidence suggests that this simplified description may not be sufficient to describe the physics of cuprate compounds~\cite{Jiang2023,Xu2024,Padma2025,Scheie2025,Li2025,lange2025hamiltonianreconstruction}. This raises the question if additional terms like next-nearest neighbor or density assisted tunneling must be considered in order to accurately describe cuprate superconductors. To address this question, it is useful to take a step back and revisit the multi-band description of cuprates in form of the Emery model. Despite recent advances with tensor networks~\cite{White2015,Ponsioen2023,Jiang2023}, Monte Carlo methods~\cite{Chiciak2020,Huang2017,mei2025magneticelectronholeasymmetrycuprates} and dynamical mean-field theory~\cite{Tseng2025}, accurate simulations of this three-band model have remained challenging for most numerical methods.

In recent years, ultracold atoms in optical lattices~\cite{Bloch2008,Gross2017} have emerged as exceptionally powerful and highly tunable quantum simulation platforms of strongly correlated systems. One of the broader goals of this field has long been to emulate the physics of real materials, and major milestones have been reached in recent years, including the achievement of ultralow temperatures~\cite{Xu2025}, spin-resolved detection~\cite{bollSpinDensityresolvedMicroscopy2016,hilkerRevealingHiddenAntiferromagnetic2017,koepsellImagingMagneticPolarons2019,Koepsell2020,Koepsell2021},, and the realization of multi-band lattice models like the Lieb lattice~\cite{lebrat2025ferrimagnetismultracoldfermionsmultiband}. 

Here, we take a further step toward this goal by showing how the three-band Emery model can be simulated in optical lattice experiments, within parameter regimes relevant to cuprates and beyond. To achieve this, we start from the Lieb lattice geometry shown in Fig.~\ref{fig:1}, which was previously realized experimentally~\cite{lebrat2025ferrimagnetismultracoldfermionsmultiband}, and extend it by introducing a potential offset on the $p$ sites -- the charge-transfer energy $\Delta_{pd}$ (see Fig.~\ref{fig:1}a,b). This offset $\Delta_{pd}$ is a key ingredient for studying the Emery model in regimes relevant to a range of different materials: $(i)$ Cuprates, where $\Delta_{pd}$ represents a key parameter for the transition temperatures \cite{Weber_2012,Ruan2016Relationshipbetweentheparentchargetransfergap,Wang2023Correlatingthecharge-transfergaptothemaximumtransitiontemperature,Yee2014charge-transferenergyinhole-dopedcuprates}. $(ii)$ Infinite-layer nickelate superconductors~\cite{Wang2024ExperimentalProgressinSuperconductingNickelates},  where $\Delta_{pd}$ is even larger than for cuprates~\cite{Botana2020SimilaritiesandDifferences,Lin2021StrongSuperexchange,Chen2022Electronicstructureofsuperconductingnickelates}; and $(iii)$ other phenomena like altermagnetism~\cite{kaushal2024altermagnetismmodifiedlieblattice}. We will show that in the setup that we propose, $\Delta_{pd}$ can be controlled via the interference of two laser beams with tunable polarization angle.  We demonstrate that this setup provides a new perspective on the large-scale physics of the Emery model using state-of-the-art quantum simulators, granting direct access to key phenomena such as Zhang-Rice singlet formation and the doping dependence of magnetic correlations. Similar regimes could also be achieved using a slightly more complex setup that involves two laser beams with different wavelengths as proposed in  Refs.~\cite{Flannigan_2021,Liberto2016}. Our work is relevant not only from an experimental standpoint but also from a theoretical perspective, as it provides a new approach to determining the physical properties of the multiband model and to identifying the key ingredients for an effective single-band description.
\\

\textit{Model.--} We start from the typical band structure of cuprates and nickelates without doping~\cite{Wang2023Correlatingthecharge-transfergaptothemaximumtransitiontemperature}, sketched in Fig.~\ref{fig:1}a,b. In the particle description (with energy $E_\mathrm{p}$), this corresponds to the (weakly interacting)  oxygen $p$-bands being fully-filled and copper $d$-bands being half-filled, with interaction energies $U_p<U_d$. With a particle-hole transformation (reversing the sign of the energy axis $E_\mathrm{p}\to E_\mathrm{h}$), this can be mapped to a system with half-filled $d$-bands and empty $p$-bands. For nickelates, the energy offset $\Delta_{pd}$ is much larger than for cuprates, leading to different physics upon hole-doping, as will become apparent later.

In the copper-oxide layers of cuprates, formed by a Lieb lattice geometry with the respective $d$-, $p_x$, $p_y$-orbitals on each unit cell, see Fig.~\ref{fig:1}c, hopping amplitudes are given by the wave function overlaps of the $p$- and $d$-orbitals. We choose a gauge of the $p$-orbitals in which all hopping amplitudes take the same sign throughout the lattice, see inset of Fig.~\ref{fig:1}c and Supplemental Material (SM)~\cite{SM}, resulting in the following version of the Emery model on the Lieb lattice
\begin{align}
\hat{\mathcal{H}} = & -t_{pd} \sum_{\langle \mathbf{i}\mathbf{j} \rangle ,\sigma} (\hat{d}^{\dagger}_{\mathbf{i}\sigma} \hat{p}_{\mathbf{j}\sigma} + \text{h.c.}) 
- \sum_{\mathbf{j}\mathbf{j^\prime} , \sigma} t_{pp}^{\mathbf{jj^\prime}} (\hat{p}^{\dagger}_{\mathbf{j}\sigma} \hat{p}_{\mathbf{j^\prime}\sigma} + \text{h.c.})\notag \\
&+ \Delta_{pd} \sum_{\mathbf{j},\sigma} \hat{p}^\dagger_{\mathbf{j}\sigma} \hat{p}_{\mathbf{j}\sigma}
+ \sum_{\nu \in \{p,d\}} U_\nu \sum_\mathbf{i} \hat{n}^\nu_{\mathbf{i}\uparrow} \hat{n}^\nu_{\mathbf{i}\downarrow}. \label{eq:3bandEmery} 
\end{align}
Here, $\hat{d}^{(\dagger)}_{\mathbf{i}\sigma}$ and $\hat{p}^{(\dagger)}_{\mathbf{i}\sigma}$ annihilate (create) holes at $d$-sites and $p=p_x,p_y$-sites respectively, and hole densities are given by \( \hat{n}^d_{\mathbf{i}\sigma} = \hat{d}^\dagger_{\mathbf{i}\sigma} \hat{d}_{\mathbf{i}\sigma} \), \( \hat{n}^p_{\mathbf{i}\sigma} = \hat{p}^\dagger_{\mathbf{i}\sigma} \hat{p}_{\mathbf{i}\sigma} \). Nearest-neighboring $d$-$p$ sites are indicated by $\langle \dots \rangle$. We consider $t_{pp}^\mathbf{ij}=t_{pp}^{(\prime)}$ for (next-)nearest-neighbor $p$-$p$ hopping, and $t_{pp}^\mathbf{ij}=0$ otherwise. The charge-transfer energy is denoted by \( \Delta_{pd} =\epsilon_p-\epsilon_d \); \( U_{d(p)} \) are the on-site Hubbard interactions for $d$ ($p$) sites. Previously used and estimated realistic choices of parameters relevant to cuprates and nickelates are $t_{pp}/t_{pd}=0,\dots,0.5$; $t_{pp}^\prime/t_{pd}=0,\dots,(t_{pp}/t_{pd})$; $U_d/t_{pd}=4.5,\dots,9.2$; and $U_p/t_{pd}=0,\dots,5.7$ and  $\Delta_{pd}/U_{d}=0.3,\dots,0.5$ (cuprates) or $\Delta_{pd}/U_d=1,\dots,1.5$ (nickelates) \cite{Hybertsen1989,Martin1996,Hanke2010,Cui2020,McMahan1988,White2015,Kent2008,MDopf1990,Jiang2023,Kowalski2021Oxygenholecontent,Tseng2025,Jiang2020criticalNatureoftheNiSpinState,Botana2020SimilaritiesandDifferences,Lin2021StrongSuperexchange,Chen2022Electronicstructureofsuperconductingnickelates,Wang2023Correlatingthecharge-transfergaptothemaximumtransitiontemperature}. In some works, also a repulsive term between $p$ and $d$ sites is considered \cite{Hanke2010,Cui2020,SM}, which however is often neglected while increasing $\Delta_{pd}$ slightly \cite{Jiang2023}. In this work, we will show that parameters in these regimes are accessible with cold atom experiments. Beyond that, the Emery model features many other interesting physical regimes, including flat-band physics, fractionalization and altermagnetism \cite{Nikoalenko2025cantedmagnetism,kaushal2024altermagnetismmodifiedlieblattice}. \\

\begin{figure*}[t]
\vspace{-0.3cm}\includegraphics[width=0.9\textwidth]{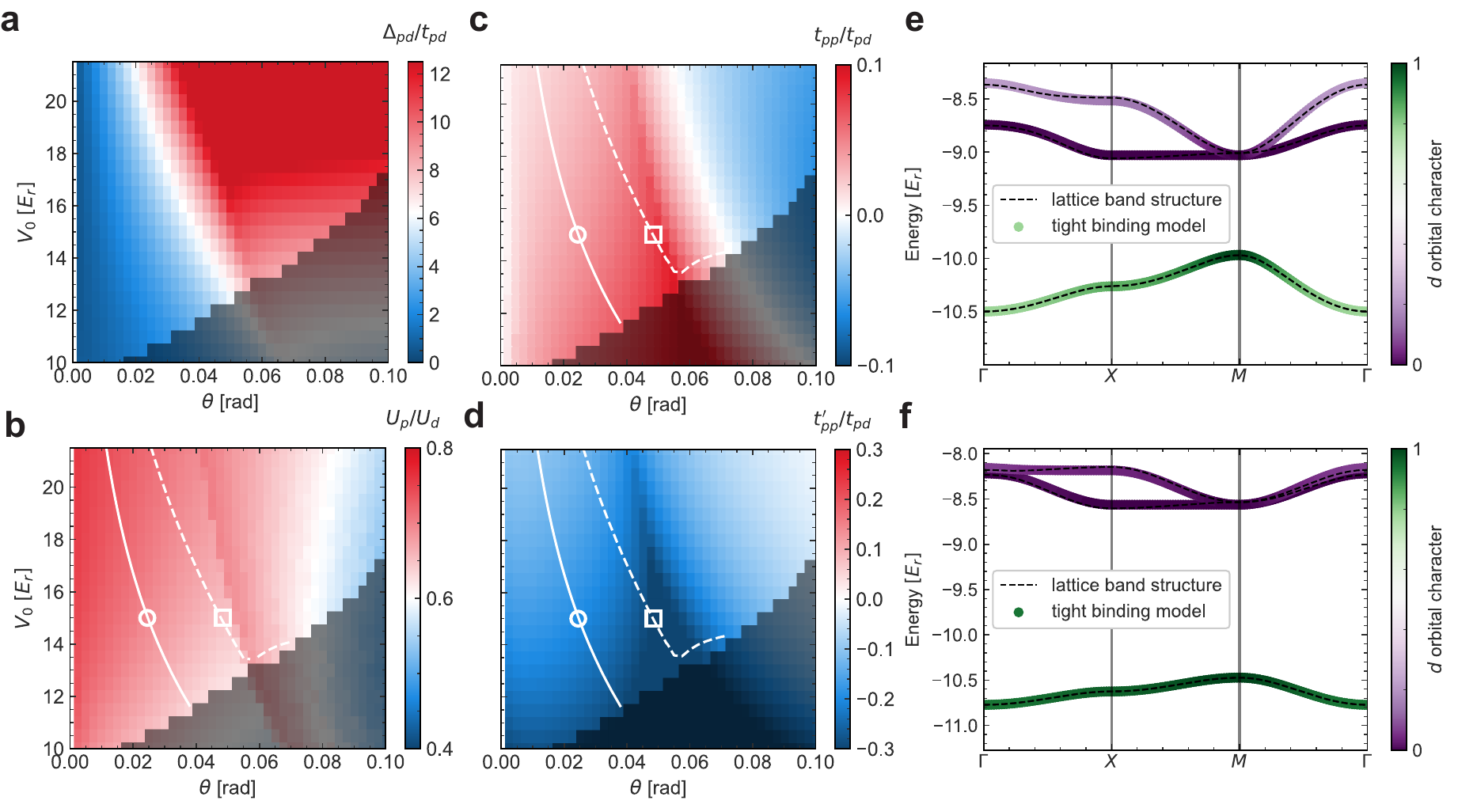}
    \caption{Accessible Emery-model parameters in the proposed optical lattice setup: $\Delta/t_{\rm pd}$ (\textbf{a}), $U_p/U_d$ (\textbf{b}), $t_{pp}/t_{pd}$ (\textbf{c}) and $t_{pp}^\prime/t_{pd}$ (\textbf{d}). The gray masks in the bottom-right corners denote the region that fourth band is too small to support an accurate three-band tight-binding description. The sharp feature appearing in \textbf{b} around $\theta\sim\SI{0.05}{rad}$ originate from a change in the band structure where the $d$ and $p$ bands begin to decouple and coincides with the trends of the other parameters~\cite{note_band}. In \textbf{b}, \textbf{c} and \textbf{d}, the two white curves indicate $\Delta_{pd}/t_{pd}=3.5$ relevant to cuprates (solid) and $\Delta_{pd}/t_{pd}=9$ relevant to nickelates (dashed). Panels \textbf{e} and \textbf{f} show band dispersions for $V_0/E_r=15$ at $\theta=\SI{0.025}{rad}$ and $\theta=\SI{0.049}{rad}$ (marked by the white circle and square in panels \textbf{c} and \textbf{d}), respectively. For panel \textbf{c} (cuprates), $\Delta_{pd}/t_{pd}=3.5$ with $t_{pd}=0.31\,E_r$, $t_{pp}/t_{pd} = 0.03$ and $t_{pp}^\prime/t_{pd} = -0.2$; for panel \textbf{d} (nickelates), $\Delta_{pd}/t_{pd}=9$ with $t_{pd}=0.24\,E_r$, $t_{pp}/t_{pd} = 0.06$ and $t_{pp}^\prime/t_{pd} = -0.3$. Dashed lines show the exact dispersions from the experimental lattice potential, while the colored lines show the reconstructed dispersions from the tight-binding Emery model, with color indicating the $d$-orbital character.}
    \label{fig:Experiment}
\end{figure*}

In the remainder of this paper, we will use the particle-hole transformed description (hole picture), and propose to simulate these holes by fermionic atoms in an optical lattice. This description is chosen because in the regime of low to intermediate dopings it translates to a small filling of holes, $\langle \hat{n}\rangle=\frac{1}{3L_xL_y}\sum_{\mathbf{i}\nu}\langle \hat{n}^\nu_\mathbf{i}\rangle=1/3 $, instead of a particle filling of $\langle \hat{n}^{(p)}\rangle=5/3$, which is numerically and experimentally less demanding. Throughout this paper, \textit{occupied} refers to occupation in the hole picture. At half-filling, holes predominantly occupy the $d$-sites, since $p$-site occupations are penalized by the energy gap $0<\Delta_{pd}<U_d$. The system can be doped by adding (removing) holes, corresponding to $\langle \hat{n}\rangle>1/3$ for hole-doping ($\langle \hat{n}\rangle<1/3$ for particle doping). The behavior upon hole-doping crucially depends on $\Delta_{pd}$: For $\Delta_{pd}\gtrsim U_d$ (nickelates), there is a large probability for additional dopants to reside on $d$-sites, whereas for $\Delta_{pd}<U_d$ (cuprates), they predominantly sit on the $p$-sites. In the limit $t_{pd}\ll \Delta_{pd}<U_d$ and setting $t_{pp}^{(\prime)}=0$ for now, a perturbative treatment reveals an antiferromagnetic (AFM) spin exchange between $d$-sites and neighboring $p$-sites~\cite{MDopf1990,Zhang1988}. This leads to the formation of the famous Zhang-Rice singlets for cuprates between one $d$-site and the symmetric oxygen superposition, 
thereby allowing to map the corresponding Wannier functions centered at the $d$-sites to a single-band $t$-$J$ model \cite{Zhang1988}. 
For the parameter regimes of cuprates, the formation of these Zhang-Rice singlets is energetically separated by an energy gap $\Delta_{\mathrm{ZR}} \gtrsim t_{pd}$, see SM~\cite{SM}. Furthermore, an AFM super-exchange coupling $J_{dd}\propto t_{pd}^4/(\Delta_{pd}^2U_d)$ between neighboring $d$-sites can be obtained in this limit  \cite{Zhang1988}. Hence, AFM ordering within the copper sites is expected for both cuprates and nickelates, but with a faster decay for nickelates at finite temperatures or upon doping, when $J_{dd}$ competes with other energy scales~\cite{Hayward2003Synthesisoftheinfinitelayer,Jiang2020criticalNatureoftheNiSpinState,Liu2020Electronicandmagneticstructureofinfinite-layerNdNiO2,Ortiz2022Magneticcorrelationsininfinite-layernickelates}.

In the following, we will explain how the the Emery model within the relevant Hamiltonian parameters can be realized in cold atom experiments in order to study the physics of cuprates and nickelates. \\

\textit{Optical lattice setup.--} As shown in Fig.~\ref{fig:1}c, geometrically, the copper-oxide planes described by the Emery model are equivalent to a Lieb lattice, as e.g. realized experimentally in cold atoms recently~\cite{lebrat2025ferrimagnetismultracoldfermionsmultiband}, but with an additional charge-transfer energy $\Delta_{pd}$ that plays the role of an energy offset between the central copper ($d$) sites and the surrounding oxygen ($p$) sites. While $\Delta_{pd}\neq 0$ could be realized by Digital Micromirror Devices (DMDs) in the setup of Ref.~\cite{weiObservationBraneParity2023,lebrat2025ferrimagnetismultracoldfermionsmultiband}, in the following we propose an alternative, passively phase-stable setup with tunable $\Delta_{pd}$. Starting from a single-beam, passively phase-stable square lattice in a bow-tie geometry~\cite{sebby-strableyLatticeDoubleWells2006,brownSpinimbalance2DFermiHubbard2017,weiObservationBraneParity2023}, we introduce this energy offset $\Delta_{pd}$ between the $d$ and $p$ sites by imprinting a polarization difference between the two passes of the lattice beam, as shown in Fig.~\ref{fig:1}d. We choose the polarization of the beam propagating along the $x$ axis to be perpendicular to the lattice plane, denoted by $V$. By inserting a half-wave plate after the first pass through the atoms, we tune the polarization of the beam on the $y$ path. Because the retro-reflected beam traverses the wave plate a second time, the polarization on the retro-reflected $x$ path remains $V$. Let $\theta$ denote the polarization angle of the $y$ path with respect to the lattice plane ($\theta=0$ and $\pi/2$ correspond to pure $H$ and $V$ polarization respectively). The interference contrast between the two axes is then proportional to $\sin\theta$, and the resulting optical-lattice potential can be written as:
\begin{equation}
\begin{split}
        V(x,y)/V_0 =& -\frac{1}{2}\qty(\cos^2(k_Lx) + \cos^2(k_Ly))\\
        &-\sin\theta\cos(k_Lx)\cos(k_Ly)
        \label{eq:Opticalpot}
\end{split}
\end{equation}
where $k_L=2\pi/\lambda_L$ is the lattice wave vector for laser wavelength $\lambda_L$; $V_0$ controls the lattice depth. By tuning the angle $\theta$, we can control the energy offset $\Delta_{pd}$ between $p$ sites and $d$ sites. We also note that an alternative approach to realizing such a lattice potential is to use three lattice beams with different polarizations whose phases are actively locked~\cite{tarruellCreatingMovingMerging2012,Xu2025}. On top of this square lattice with offset between nearest neighboring sites, we propose to superimpose a patterned array of repulsive potentials as in Ref.~\cite{lebrat2025ferrimagnetismultracoldfermionsmultiband}, selectively lifting the energy of every fourth ''copper" site using a high-resolution microscope projection system to achieve the Lieb lattice geometry~\cite{weiObservationBraneParity2023,lebrat2025ferrimagnetismultracoldfermionsmultiband}.

By fitting the single-particle band structure of the Emery model, Eq.~\eqref{eq:3bandEmery}, to the optical-lattice band structure, we extract the effective charge-transfer gap and hopping parameters shown in Fig.~\ref{fig:Experiment}a,c,d. The regime with sizable positive offsets, $\Delta_{pd}/t_{pd}\approx 2,\dots,8$, coincides with nearest-neighbor hopping $t_{pp}/t_{pd}\approx 0.1$ and next-nearest-neighbor hopping $t_{pp}^\prime/t_{pd}\approx -0.2$. While the latter is larger in magnitude than values typically used for cuprates, we show in the SM that the relevant observables change only weakly compared to the canonical cuprate parameter set~\cite{SM}.
The accessible interaction strengths are on the order of $U_p/U_d\approx 0.6$, relevant to cuprates (see Fig.~\ref{fig:Experiment}b and SM~\cite{SM}), with the overall scale tunable via a Feshbach resonance~\cite{chinFeshbachResonancesUltracold2010}. 
In the cuprate regime, Fig.~\ref{fig:Experiment}e shows that the fitted band structure for a representative angle $\theta=\SI{0.025}{rad}$ ($\SI{1.4}{deg}$), corresponding to $\Delta_{pd}/t_{pd}=3.5$, agrees well with the full band structure of this model. Here, lattice depth is expressed in units of the recoil energy $E_r=h^2/(2ma^2)$, with atomic mass $m$ and lattice constant $a$ equaling $\lambda_L$ in this setup (see Fig.~\ref{fig:1}d,e). For reference, using $^{6}$Li atoms and a $\lambda_L=\SI{1064}{nm}$ laser, the recoil energy is $E_r=h\times\SI{7.3}{kHz}$. With these parameters, the inter-orbital hopping is $t_{pd}=0.31\,E_r=h\times\SI{2.26}{kHz}$. This relatively large energy scale makes the quantum simulation compatible with current quantum gas microscopes and their achievable experimental timescales~\cite{grossQuantumGasMicroscopy2021}.
By increasing $\theta$ at the same lattice depth, we can continuously increase the relative charge-transfer gap $\Delta_{pd}/t_{pd}$ and thereby access the nickelate regime; for example, $\Delta_{pd}/t_{pd}=9$ at $\theta=\SI{0.049}{rad}$ ($\SI{2.8}{deg}$), as shown in Fig.~\ref{fig:Experiment}f. Achieving polarization control precision and purity below $\SI{1}{deg}$ is feasible with standard high-quality optical components and careful calibration~\cite{Wen:21, robensLowEntropyStatesNeutral2017}.
Note that in the band structure momenta are shifted relative to Emery-model simulations that do not include the gauge transformation~\cite{Maier2013,Tseng2025} by $\mathbf{k}\to \mathbf{k}+(\pi,\pi)$, see SM~\cite{SM}.
To access the low-temperature regime, we propose using cooling schemes based on entropy redistribution similar to Ref.~\cite{Xu2025}.

\begin{figure}[t]   \includegraphics[width=0.49\textwidth]{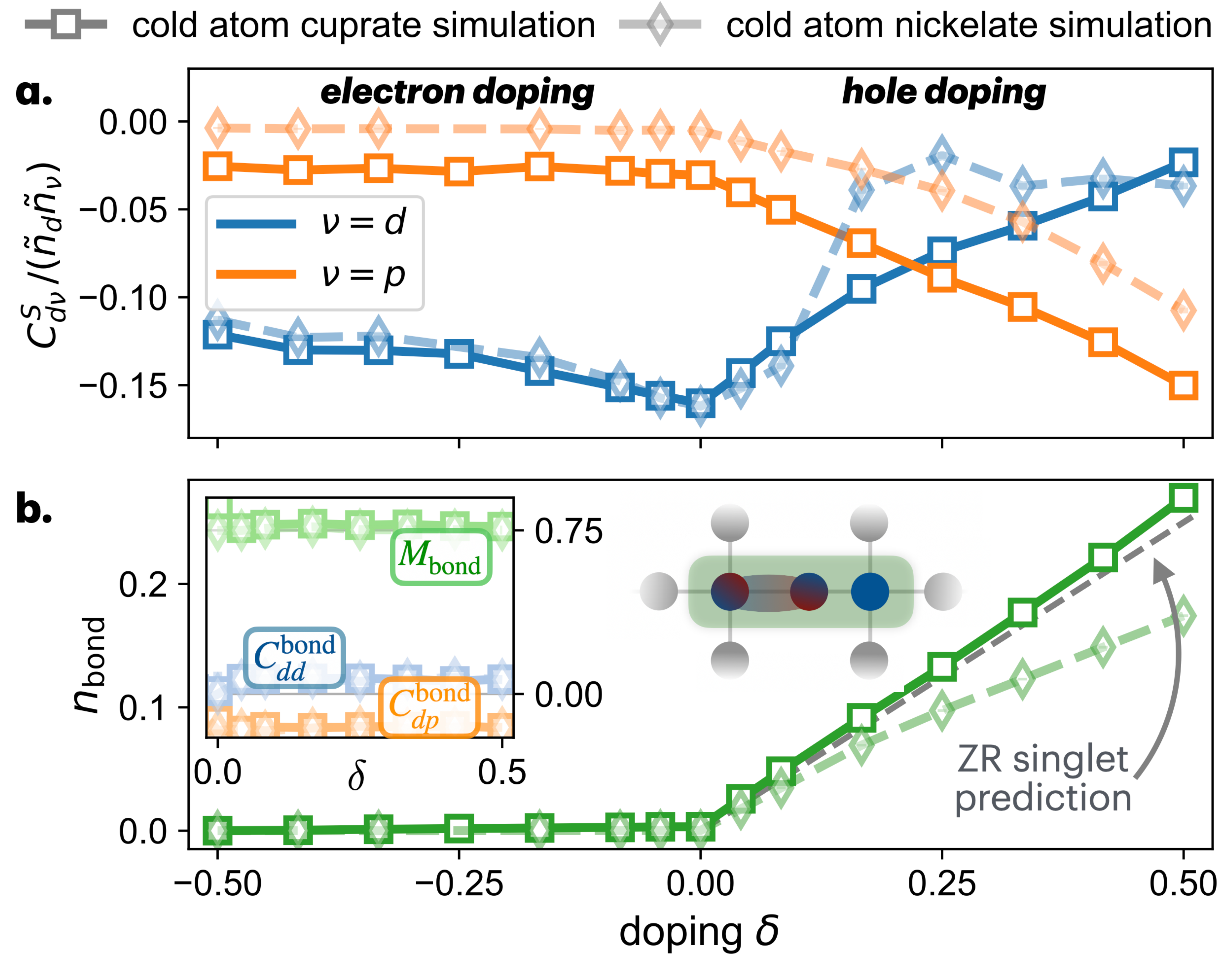}
    \caption{Numerical results for a cylinder with $12\times 2$ unit cells ($72$ sites) for two sets of parameters relevant to the cuprates ($\Delta/t_{pd}=3.5$) and infinite-layer nickelates ($\Delta/t_{pd}=9.0$) that are accessible for quantum simulation platforms and changing the doping $\delta=N_\mathrm{h}/(L_xL_y)$. \textbf{a)} Average nearest-neighbor spin correlations $C^S_{d\nu}$ divided by average densities $\tilde{n}_\nu$, Eq.~\eqref{eq:SS}, with $\nu=d (p)$ in blue (orange).  \textbf{b)} Number of occupied $d$-$p$-$d$ bonds (see sketch). For Zhang-Rice singlets, we expect $N_\mathrm{bond}=N_\mathrm{h}/2$ (gray line), see SM~\cite{SM}. The inset shows the spin correlations $C^\mathrm{bond}_{d\nu}$ (blue and orange) and the total spin $M_\mathrm{bond}$ (green), both post-selected for occupied $d$-$p$-$d$ bonds. For all calculations, we set $U_d/t_{pd}=8.0$, $U_p/t_{pd}=3.0$, $t_{pp}/t_{pd}=0.1$ and $t_{pp}^\prime/t_{pd}=-0.2$. All observables are evaluated from $2000$ snapshots drawn from the MPS, errorbars denoting the error of the mean are smaller than the markers. }
    \label{fig:DMRG}
\end{figure}

In the following, we will demonstrate that the parameters accessible with our proposed experiment allow one to observe key phenomena like Zhang-Rice singlet formation or the asymmetry of the magnetic state at hole and particle doping. Furthermore, we will show how $\Delta_{pd}$ allows to interpolate between the clearly Zhang-Rice dominated cuprate regime to a Hubbard-Mott regime relevant to infinite-layer nickelates.  \\

\textit{Numerical results.--} The setup described above allows one to access the physics of large-scale Emery models beyond the restrictions of numerical methods. Here, we nevertheless use numerical simulations for cylinders with $12\times 2$ unit cells accessible to DMRG calculations in order to $(i)$ predict possible observables for the proposed experiment and $(ii)$ compare the results for different choices of Hamiltonian parameters. Considering the experimentally accessible parameters $U_d/t_{pd}=8.0$, $U_p/t_{pd}=3.0$, $t_{pp}/t_{pd}=0.1$ and $t_{pp}^\prime/t_{pd}=-0.2$, we vary the charge-transfer energy from $\Delta_{pd}/t_{pd}=3.5$ relevant to cuprates \cite{Hanke2010,Cui2020,McMahan1988,Jiang2023,SM} to $\Delta_{pd}/t_{pd}=9.0$ relevant to infinite-layer nickelates \cite{Wang2024ExperimentalProgressinSuperconductingNickelates}. All calculations are performed using SyTen~\cite{syten1,syten2}, exploiting the $U(1)\times U(1)$ symmetry corresponding to the conservation of the particle number and magnetization $S_z$, see SM \cite{SM}.

Fig.~\ref{fig:DMRG} shows observables that are evaluated from $N_S$ spin-projected snapshots, $s=s_1,s_2\dots s_{N}$ with $N=3L_xL_y$ and $s_i=\{-1/2,+1/2,0\}$, that are drawn from the MPS, similar to the data that would be available in an experiment with full spin resolution~\cite{Koepsell2020}. We consider both hole and electron dopings, which corresponds to the addition or removal of $\pm N_\mathrm{h}$ holes from the $\langle n\rangle=1/3$ filled state, $\delta = N_\mathrm{h}/(L_xL_y)$. In Fig.~\ref{fig:DMRG}a, we evaluate the nearest-neighbor spin correlations 
\begin{align}
     C^S_{d\nu} = \frac{1}{L_xL_y}\sum_s \sum_{\langle\mathbf{i} \mathbf{j}\rangle_{d \nu}} s_\mathbf{i}\cdot s_\mathbf{j},
    \label{eq:SS}
\end{align}
where $\langle\mathbf{i} \mathbf{j}\rangle_{d \nu}$ involves a $d$-site and a closest $\nu\in \{p,d\}$-site. We divide by the average spin-densities (where doublons and empty sites both contribute $s_\mathbf{i}=0$), $
 \tilde{n}_\nu = \frac{1}{3L_xL_y}\sum_{\mathbf{i}\in \nu\mathrm{-sites}} \vert 2 s_\mathbf{i}\vert $. 
 As expected from the AFM (super)exchange in the $t_{pd}\to 0$ limit, both $C_{dd}^S/(\tilde{n}_d^2)$ and $C_{dp}^S/(\tilde{n}_d\tilde{n}_p)$ are negative, in agreement with previous numerical results~\cite{Ponsioen2023,Jiang2023,mei2025magneticelectronholeasymmetrycuprates}. The correlations $C_{dd}^S/(\tilde{n}_d^2)$ decrease with doping $\delta$, both for electron and particle doping. This is very similar to the single-band Fermi-Hubbard model on the square lattice, for which spin correlations also decrease with doping~\cite{Koepsell2021}. Note however that the spin correlations $C_{dd}^S/\tilde{n}_d$ considered here are not exactly the same as the single-band particle-spin counterpart, as a rigorous downfolding to a single-band description would (besides a particle-hole transformation) involve Wannier states centered at the Cu-sites and not the bare Cu-sites \cite{Jiang2023}. 
 
 Furthermore, the asymmetry between particle and hole doped sides is not captured by single-band $t$-$J$ and Fermi-Hubbard models without next-neighbor tunneling but is clearly visible here. It is especially pronounced for $\Delta_{pd}>U_d$ relevant to nickelates where the suppression of AFM order is expected~\cite{Hayward2003Synthesisoftheinfinitelayer,Jiang2020criticalNatureoftheNiSpinState,Liu2020Electronicandmagneticstructureofinfinite-layerNdNiO2,Ortiz2022Magneticcorrelationsininfinite-layernickelates}.
The spin-correlations $C_{dp}^S/ (\tilde{n}_d\tilde{n}_p)$ between neighboring $d$ and $p$ sites remain vanishingly small for the full range of considered electron-dopings, while for hole-doping they increase linearly with $\delta$, with a larger absolute value for the cuprate regime. This is a first indication for Zhang-Rice singlet formation for cuprates, that is only expected in the hole-doped regime, when additional holes delocalize on the $p$-sites.\\

To obtain further insights into the formation of Zhang-Rice singlets, we consider occupied $d$-$p$-$d$ bonds (represented by $\vert_{\mathbf{ijk}={\mathrm{occ.}}}$ in the following). The number of such occupied bonds $N_\mathrm{bond}$ normalized to the number of unit cells, $n_\mathrm{bond}=N_\mathrm{bond}/(L_xL_y)$, vanishes in the electron-doped regime and is proportional to the dopant density $n_\mathrm{h}=N_\mathrm{h}/(L_xL_y)$ in the hole-doped regime. Specifically, we observe $n_\mathrm{bond}=n_\mathrm{h}/2$ (see gray dashed line in Fig.~\ref{fig:DMRG}b) for the hole-doped cuprate regime, which is expected for Zhang-Rice singlet formation, see SM~\cite{SM}. For the nickelate regime, where the large charge-transfer energy suppresses Zhang-Rice singlet formation~\cite{Jiang2020criticalNatureoftheNiSpinState}, $n_\mathrm{bond}$ is significantly smaller.  

For the occupied bonds in both cuprate and nickelate regimes, we calculate the nearest-neighbor spin correlations, $ C^S_{d\nu}\vert_{\mathbf{ijk}={\mathrm{occ.}}}$, as well as the absolute bond magnetization, 
\begin{align}
    M_{\mathrm{bond}}=\frac{3}{\Tilde{N}}\sum_s \sum_{
\mathbf{i}\mathbf{j}\mathbf{k}\in (d\text{-}p\text{-}d)}(s_\mathbf{i} + s_{\mathbf{j}}+s_{\mathbf{k} } )^2\Big|_{ \mathbf{i}\mathbf{j}\mathbf{k}=\mathrm{occ.}},
    \label{eq:Mbond}
\end{align}
where $\Tilde{N}=N_\mathrm{bond}N_S$ and the factor of three takes into account the SU(2) symmetry of the system, see SM \cite{SM}. In the inset of Fig.~\ref{fig:DMRG}b, we observe an almost constant value of $M_\mathrm{bond}$ close to the minimal possible value of $0.75$ throughout all considered hole dopings $\delta>0$, with the value of $0.75$ indicating the formation of a singlet on the rung (contributing with $0$ to $M_\mathrm{bond}$), and the third spin giving rise to the value of $0.75$. For infinite temperature, when all possible configurations of occupied bonds occur, $M_\mathrm{bond}\to3(6\cdot 0.5^2 +2\cdot 1.5^2 )/8=2.25$ would be significantly larger. Together with the observation that the postselected $d$-spin correlations $C^\mathrm{bond}_{dd}:=C^S_{dd}\vert_{\mathbf{ijk}={\mathrm{occ.}}}$ are slightly ferromagnetic while $C^\mathrm{bond}_{dp}:=C^S_{dp}\vert_{\mathbf{ijk}={\mathrm{occ.}}} \approx -0.15$ is AFM, we conclude that the singlet is predominantly formed between $p$- and $d$-sites, as expected for Zhang-Rice singlets.\\

\textit{Conclusion and outlook.--} We have proposed an experimentally robust scheme to realize the three-band Emery model in state-of-the-art ultracold atom quantum simulators. By combining a Lieb-lattice geometry with a tunable charge-transfer energy gap between $p$- and $d$-sites, our approach enables access to parameter regimes relevant to cuprate and infinite-layer nickelate superconductors. Using DMRG calculations as a benchmark, we demonstrated that hallmarks of cuprate physics -- most notably Zhang–Rice singlet formation and a particle-hole asymmetry in magnetic correlations -- are robustly observable within the experimentally accessible parameter window.
Our work hence establishes cold atom platforms as a powerful route to studying multiband strongly correlated models on system sizes beyond current numerical limits. Experimental realizations of our proposal are of relevance from two perspectives: First, they allow the study of cuprate or nickelate physics within a model that is closer to the real materials than single-band models, including the emergence of quasiparticles like magnetic polarons, the formation of Zhang-Rice singlets with analogies to cuprate experiments~\cite{ye2023visualizingzhangricesingletmolecular,Ye2013}, or superconductivity in the three-band model. Second, it enables a systematic investigation on how multi-band physics is translated to effective interaction terms in single-band models, by applying downfolding schemes based on the measurement of single-particle coherences~\cite{Gluza2021,Denzler2024,kiser2025gaussiantomographycoldatomsimulators} and consecutively the construction of \textit{many-body} Wannier orbitals, similar to the numerical work of Ref.~\cite{Jiang2023}. More broadly, our results pave the way toward simulations of real-material Hamiltonians and a deeper microscopic understanding of unconventional superconductivity using cold atom platforms.\\

\textit{Note.--} During the completion of this work, we became aware of a related proposal that considers the implementation of the Emery model using a bichromatic optical lattice~\cite{McCabe}.\\

\textit{Acknowledgments.--} 
We wish to thank Xiaoxiao Wang for valuable discussions. This research was supported by the Max Planck Society (MPG), the German Federal Ministry of Research, Technology and Space (BMFTR grant agreement 13N15890, FermiQP), the Deutsche
Forschungsgemeinschaft (DFG, German Research Foundation)
under Germany’s Excellence Strategy EXC-2111 Grant
No. 390814868 and the European Research Council
(ERC) under the European Union’s Horizon 2020 research
and innovation program (Grant Agreement No. 948141),
ERC Starting Grant SimUcQuam, ERC StartingGrant UniRand (Grant No. 94824) and ERC Starting Grant QuaQuaMA (Grant No. 101217531). Numerical simulations were performed on the Arnold Sommerfeld Cluster. DMRG calculations have been done using the \textsc{SyTen} toolkit, developed and maintained by C. Hubig, F. Lachenmaier, N.-O. Linden, T. Reinhard, L. Stenzel, A. Swoboda, M. Grunder, S. Mardazad, F. Pauw and S. Paeckel. Information is available at \href{https://syten.eu/}{www.syten.eu}.





\bibliography{references.bib}
\widetext


\newpage
\pagebreak
\appendix

\newpage

\newpage
\pagebreak
\appendix
\widetext

\newpage

\begin{center}
\textbf{\large Supplementary Materials}
\end{center}
\setcounter{equation}{0}
\setcounter{figure}{0}
\setcounter{table}{0}
\setcounter{page}{1}
\makeatletter
\renewcommand{\theequation}{S\arabic{equation}}
\renewcommand{\thefigure}{S\arabic{figure}}

\section{The gauge transformation}

\begin{figure*}[b]
    \centering
    \includegraphics[width=0.95\textwidth]{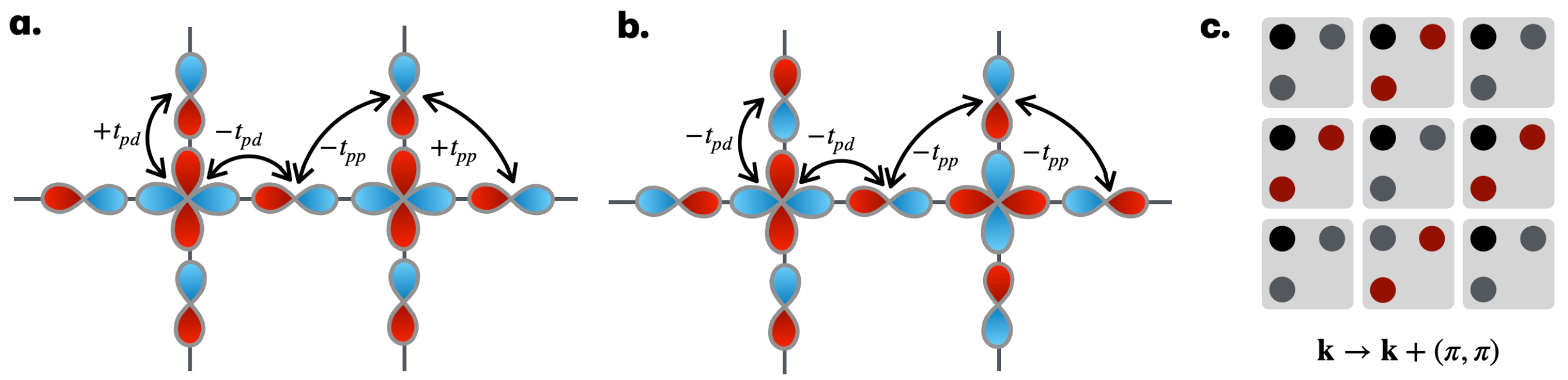}
    \caption{\textbf{a.} Copper-oxide layer orbitals with the explicit phases from the copper \(d_{x^2-y^2} \) and oxygen $p_{x,y}$ orbitals. \textbf{b.} Copper-oxide layer orbitals after the gauge transformation Eq.~\eqref{eq:trafo}. \textbf{c.} The transformation Eq. \eqref{eq:trafo} transforms the signs of the $\hat{p}_\mathbf{j} $ operators at every second unit cell, resulting in a redefinition of the quasi-momenta from $\textbf{k}\to \textbf{k}+(\pi,\pi)$. }
    \label{fig:CuOOrbs}
\end{figure*}

Without the gauge transformation employed in the main text, the copper-oxide layers of cuprates are described by the following Hamiltonian:
\begin{align}
\hat{\mathcal{H}} =&\sum_{ \mathbf{i},\mathbf{j},\sigma} V_{\mathbf{i}\mathbf{j}}^{dp}\left(\hat{d}_{\mathbf{i}\sigma}^\dagger \hat{p}_{\mathbf{j}\sigma} + \mathrm{h.c.}\right) +\sum_{ \mathbf{j},\mathbf{j^\prime},\sigma} V_{\mathbf{j}\mathbf{j^\prime}}^{pp}\left( \hat{p}_{\mathbf{j}\sigma}^\dagger \hat{p}_{\mathbf{j^\prime}\sigma} + \mathrm{h.c.} \right)
+ \Delta_{pd} \sum_{\mathbf{j},\sigma}  \hat{p}_{\mathbf{j}\sigma}^\dagger \hat{p}_{\mathbf{j}\sigma}
+ \sum_{\nu=p,d} U_\nu \sum_\mathbf{i} \hat{n}^\nu_{\mathbf{i}\uparrow}\hat{n}^\nu_{\mathbf{i}\downarrow}
\label{eq:ThreebandwithPhases}
\end{align}
Here, $\hat{p}_{\mathbf{j}\sigma}^\dagger$ and $\hat{d}_{\mathbf{i}\sigma}^\dagger$ create holes on $p_x,\,p_y$= and $d_{x^2-y^2}$-sites, respectively. \( U_{d(p)} \) is the on-site Coulomb repulsion at a $d(p)$-sites. The hybridization matrix \( V_{\mathbf{i}\mathbf{j}} \) is proportional to the wave-function overlap of $d$ and $p$ orbitals, schematically sketched in Fig.~\ref{fig:CuOOrbs}a, i.e.
\begin{equation}
V_{\mathbf{i}\mathbf{j}}^{dp} = (-1)^{M^{dp}_{\mathbf{i}\mathbf{j}}} t_{pd},\quad V_{\mathbf{j}\mathbf{j^\prime}}^{pp} = (-1)^{M^{pp}_{\mathbf{j}\mathbf{j^\prime}}} t_{pd}
\end{equation}
(where \( t_{pd(p)} \) are the amplitudes of the hybridization) for nearest $d$-$p$ neighbors and for nearest and next-nearest $p$-$p$ neighbors. Using nearest-neighbor $d$-$p$ distances as the unit length $a=1$, we have
\begin{itemize}
    \item \( M_{\mathbf{i}\mathbf{j}}^{dp} = 0 \) if \( \mathbf{j} = \mathbf{i} - \hat{x} \) or \( \mathbf{i} +  \hat{y} \) and  \( M_{\mathbf{i}\mathbf{j}}^{dp} = 1 \) if \( \mathbf{j} = \mathbf{i} +  \hat{x} \) or \( \mathbf{i} - \hat{y} \).
    \item \( M_{\mathbf{j}\mathbf{j^\prime}}^{pp} = 0 \) if \( \mathbf{j^\prime} = \mathbf{j} - \hat{x}+\hat{y} \) or \( \mathbf{j} + \hat{x}- \hat{y} \) and \( M_{\mathbf{j}\mathbf{j^\prime}}^{pp} = 1 \) if \( \mathbf{j^\prime} = \mathbf{j} +  \hat{x} +\hat{y}\) or \( \mathbf{j} - \hat{x}-\hat{y} \).
\end{itemize}
This can be seen from Fig. \ref{fig:CuOOrbs}a, with the phases of the orbitals denoted by the color. The signs of $M_{\mathbf{i}\mathbf{j}}^{dp}$, $M_{\mathbf{j}\mathbf{j^\prime}}^{pp}$ result from multiplying these phases. We furthermore consider a next-nearest neighbor $p$-$p$ hopping with $V_{\mathbf{j}\mathbf{j^\prime}}^{pp}=-t_{pp}^\prime$ and set all other, further-ranged hopping elements to zero. For more details, see Refs. \cite{Zhang1988,Kung2016}.\\

We now transform both $\hat{p}_{\mathbf{j} \sigma}^\dagger$ and $\hat{d}_{\mathbf{i} \sigma}^\dagger$ orbitals as shown in Fig. \ref{fig:CuOOrbs}b, s.t. the resulting hopping has negative sign everywhere. More precisely, this amounts to
\begin{align}
    \hat{p}_\mathbf{j}  \to -(-1)^{M_{\mathbf{i} \mathbf{j} }^{pd}}\hat{p}_\mathbf{j} ,
    \label{eq:trafo}
\end{align}
i.e. at every second unit cell we transform the signs of the $\hat{p}_\mathbf{j} $ operators, see Fig. \ref{fig:CuOOrbs}c. Under this transformation,
\begin{align}
    \sum_{\langle \mathbf{i} ,\mathbf{j}  \rangle,\sigma} V_{\mathbf{i} \mathbf{j} }^{dp}\left(\hat{d}_{\mathbf{i} \sigma}^\dagger \hat{p}_{\mathbf{j} \sigma} + \mathrm{h.c.}\right)\to -t_{pd}\sum_{\langle \mathbf{i} ,\mathbf{j}  \rangle,\sigma} (-1)^{M_{\mathbf{i} \mathbf{j} }^{dp}+M_{\mathbf{i} \mathbf{j} }^{dp}}\left(\hat{d}_{\mathbf{i} \sigma}^\dagger \hat{p}_{\mathbf{j} \sigma} + \mathrm{h.c.}\right)=-t_{pd}\sum_{\langle \mathbf{i} ,\mathbf{j}  \rangle,\sigma}\left( \hat{d}_{\mathbf{i} \sigma}^\dagger \hat{p}_{\mathbf{j} \sigma} + \mathrm{h.c.}\right)
\end{align}
and 
\begin{align}
    \sum_{\langle \langle \mathbf{j},\mathbf{j^\prime}\rangle \rangle,\sigma} V_{\mathbf{j}\mathbf{j^\prime}}^{pp}\left(\hat{p}_{\mathbf{j}\sigma}^\dagger \hat{p}_{\mathbf{j^\prime}\sigma} + \mathrm{h.c.} \right)\to t_{pp} \sum_{\langle \langle \mathbf{j},\mathbf{j^\prime}\rangle \rangle,\sigma} (-1)^{M_{\mathbf{i}\mathbf{j}}^{pd}+M_{\mathbf{i}\mathbf{j^\prime}}^{pd}+M_{\mathbf{j}\mathbf{j^\prime}}^{pp}}\left(\hat{p}_{\mathbf{j}\sigma}^\dagger \hat{p}_{\mathbf{j^\prime}\sigma} + \mathrm{h.c.} \right)=-t_{pp}\sum_{\langle \langle \mathbf{j},\mathbf{j^\prime}\rangle \rangle,\sigma} \left(\hat{p}_{\mathbf{j}\sigma}^\dagger \hat{p}_{\mathbf{j^\prime}\sigma} + \mathrm{h.c.}\right).
\end{align}
For the next-nearest neighbor $p$-$p$ hopping, the transformation results in positive overlaps $+t_{pp}^\prime$. The total transformed Hamiltonian is given by Eq. \eqref{eq:3bandfinal}. Note that this transformation will not change any observable that involves two $\hat{p}_i$ operators, like densities, density correlations or spin correlations. However, the transformation Eq. \eqref{eq:trafo} has to considered when e.g. pairing correlations involving the $p$-bands are considered. Furthermore, it leads to a redefinition of the quasi-momenta $\textbf{k}\to \textbf{k}+(\pi,\pi)$, as illustrated in Fig.~\ref{fig:CuOOrbs}c. 


\section{Comparison to the canonical set of parameters for cuprates}

\begin{figure*}[t]
    \centering
    \includegraphics[width=0.9\textwidth]{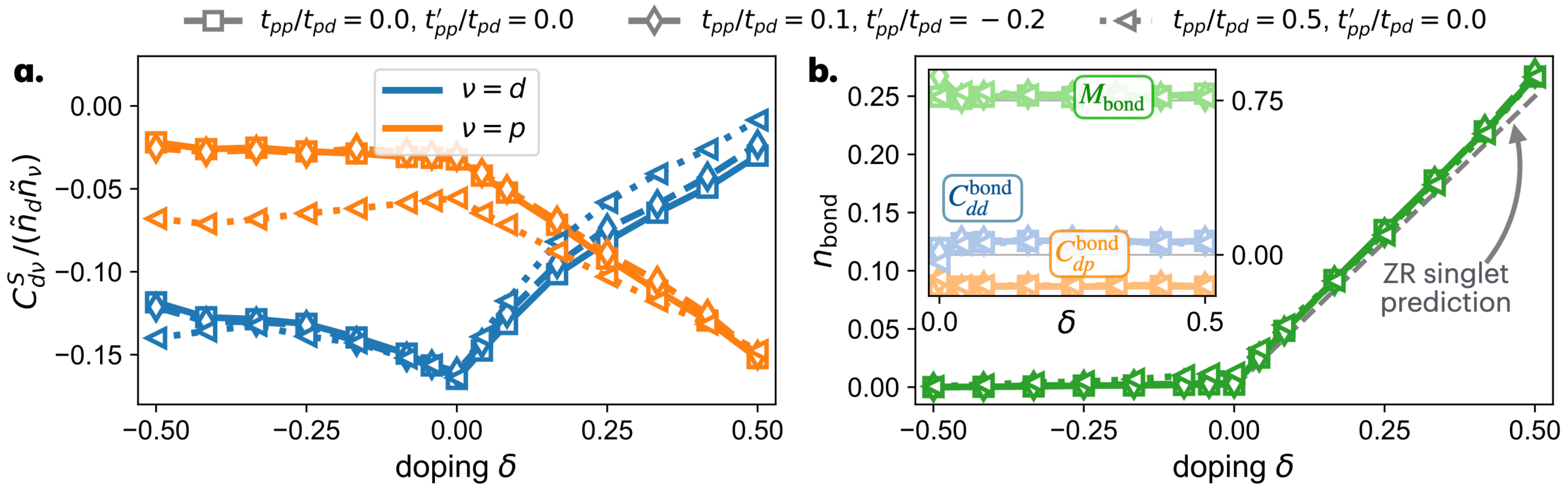}
    \caption{Numerical results for a cylinder with $12\times 2$ unit cells ($72$ sites) for two sets of parameters relevant to the cuprates ($t_{pp}/t_{pd}=0.0(0.5)$, $t_{pp}^\prime/t_{pd}=0.0$) and the experimentally accessible set ($t_{pp}/t_{pd}=0.1$, $t_{pp}^\prime/t_{pd}=-0.2$), changing the doping $\delta=N_\mathrm{h}/(L_xL_y)$. \textbf{a.} Average nearest-neighbor spin correlations $C^S_{d\nu}$, Eq.~\eqref{eq:SS}, with $\nu=d (p)$ in blue (orange) and divided by the average $\nu$-site density $\tilde{n}_\nu$ (without doublon-hole resolution).  \textbf{b.}: Number of occupied $d$-$p$-$d$ bonds (see sketch). For Zhang-Rice singlets, we expect $N_\mathrm{bond}=N_\mathrm{h}/2$ (gray line), see SM~\cite{SM}. The inset shows the spin correlations $C^\mathrm{bond}_{d\nu}$ (blue and orange) and the total spin $M_\mathrm{bond}$ (green), both post-selected for occupied $d$-$p$-$d$ bonds. For all calculations, we set $U_d/t_{pd}=8.0$, $U_p/t_{pd}=3.0$ and $\Delta_{pd}/t_{pd}=3.5$. All observables are evaluated from $2000$ snapshots drawn from the MPS, errorbars denoting the error of the mean are smaller than the markers.}
    \label{fig:DMRG2}
\end{figure*}

As discussed in the main text, the canonical Emery parameters for cuprate in the literature are $t_{pp}/t_{pd}=0,\dots,0.5$; $t_{pp}^\prime/t_{pd}=0,\dots,t_{pp}/t_{pd}$; $\Delta_{pd}/t_{pd}=2.8,\dots,3.5$; $U_d/t_{pd}=4.5,\dots,9.2$; and $U_p/t_{pd}=0,\dots,5.7$. While most parameters are accessible in our proposed experimental setup -- including $U_d$, $U_p$, and $\Delta_{pd}$ -- we encounter that the hopping terms are restricted to $t_{pp}/t_{pd}\approx 0.1$ and relatively large $t_{pp}^\prime/t_{pd}\approx -0.2$. However, Fig.~\ref{fig:DMRG2} shows that the exact choice of $t_{pp}^{\prime}$ does not change the observables that we study significantly. Increasing $t_{pp}$ significantly from $t_{pp}=0.0$ to $t_{pp}/t_{pd}=0.5$ slightly changes $C^S_{dp}/(\tilde{n}_d\tilde{n}_p)$, but leaves all other observables almost unchanged (see triangle markers).  

\section{Number of Zhang-Rice singlets in the diagonal density basis}
This section will provide an explanation why the number of $d$-$p$-$d$ bonds, $N_\mathrm{bond}$, was observed to scale as $N_\mathrm{bond}=N_\mathrm{h}/2$ in the hole-doped cuprate regime.

From the analysis in the main text, we expect that the number of bonds, $N_\mathrm{bond}$, is proportional to the number of Zhang-Rice singlets, which form for every introduced dopant $N_\mathrm{h}$. The additional factor of $1/2$ arises since $N_\mathrm{bond}$ is measured in the diagonal density basis, while the actual Zhang-Rice density is given by $\langle (\hat{\psi}^{\mathrm{ZR}}_\mathbf{i})^\dagger\hat{\psi}^{\mathrm{ZR}}_\mathbf{i}\rangle$ with $\psi^{\mathrm{ZR}}_\mathbf{i}=\frac{1}{\sqrt{2}}\left( \hat{d}_{\mathbf{i}\uparrow}\hat{P}^S_{\mathbf{i}\downarrow}-\hat{d}_{\mathbf{i}\downarrow}\hat{P}^S_{\mathbf{i}\uparrow}\right)$ and
\begin{equation}
\hat{P}_{\mathbf{i}\sigma}^{(S,A)}
=
\frac{1}{2}
\sum_{\mathbf{l}:\langle \mathbf{i},\mathbf{l}\rangle}
(\pm 1)^{M_{\mathbf{i},\mathbf{l}}}
\, \hat{p}_{\mathbf{l}\sigma},
\label{eq:P}
\end{equation}
and hence also features offdiagonal terms.

The diagonal contribution to $\langle (\hat{\psi}^{\mathrm{ZR}}_\mathbf{i})^\dagger\hat{\psi}^{\mathrm{ZR}}_\mathbf{i}\rangle$ that is probed by counting the occupied $d$-$p$-$d$ bonds is given by
\begin{align}
    \mathrm{Diag}\left[\sum_\mathbf{i}\langle (\hat{\psi}^{\mathrm{ZR}}_\mathbf{i})^\dagger\hat{\psi}^{\mathrm{ZR}}_\mathbf{i}\rangle\right] = \frac{1}{8} \sum_\mathbf{i}\langle \hat{n}_\mathbf{i}^d\sum_{\mathbf{l}:\langle \mathbf{i},\mathbf{l}\rangle}\hat{n}_\mathbf{l}^p\rangle\approx \frac{1}{8} \sum_\mathbf{i}\langle \hat{n}_\mathbf{i}^d\rangle\sum_{\mathbf{l}:\langle \mathbf{i},\mathbf{l}\rangle}\langle\hat{n}_\mathbf{l}^p\rangle\approx \frac{1}{8} \sum_\mathbf{i}1 \cdot \sum_{\mathbf{l}:\langle \mathbf{i},\mathbf{l}\rangle}\frac{N_\mathrm{h}}{L_xL_y}=\frac{N_\mathrm{h}}{2}.
\end{align}
Hereby, we have assumed that the $d$-sites are half-filled and the $N_\mathrm{h}$ additional holes reside on the $p$-sites.

\section{Estimation of Zhang-Rice energy scales}

\begin{figure*}[b]   \includegraphics[width=0.9\textwidth]{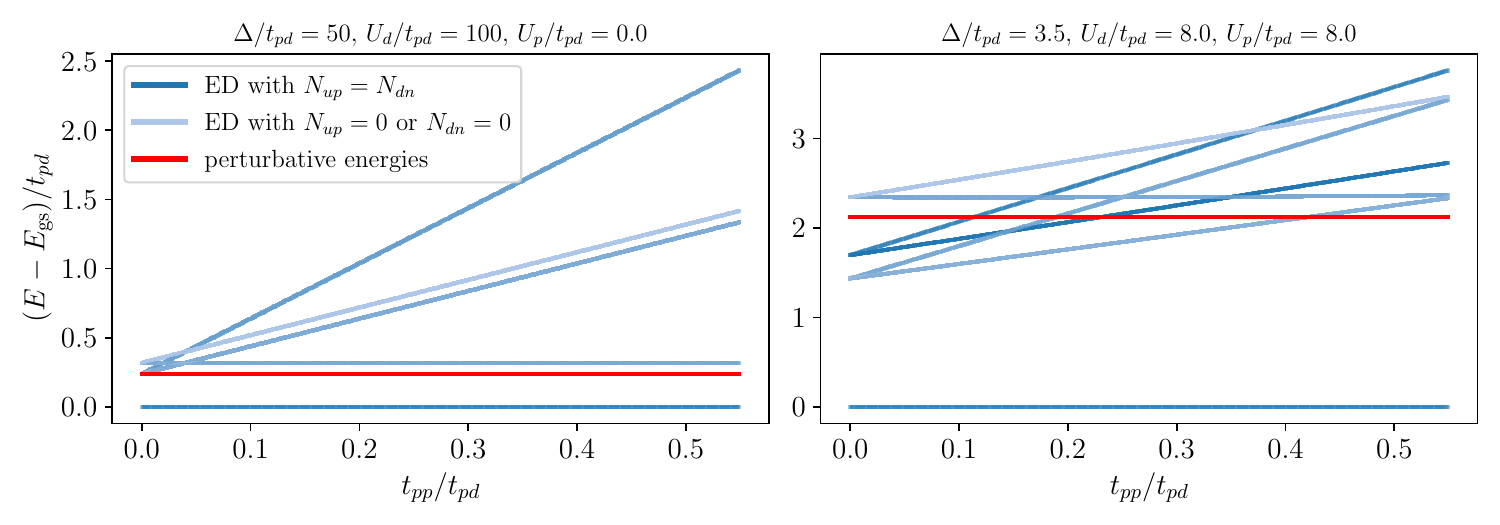}
    \caption{Exact diagonalization (ED) energies $E$ subtracted by the ground state energy $E_\mathrm{gs}$ for two holes on a plaquette with one $d$-site and four surrounding $p$-sites (blue), compared to the perturbative estimation, $E_{Ss}-E_{As}$, (red) for two different parameter choices in the two panels. }
    \label{fig:Gaps}
\end{figure*}

In the following, we show that the Zhang-Rice formation is very robust since it is protected by a large energy gap. This becomes apparent using perturbative arguments (following Ref.~\cite{Zhang1988}) or exact diagonalization of a single Zhang-Rice building block, i.e. a $d$-site surrounded by four $p$-sites. 

In order to define Zhang-Rice singlets, we follow Ref.~\cite{Zhang1988} and consider the four surrounding $p$-sites $\{\mathbf{i}\} $ around a $p$-site. Their superposition is given by Eq.~\eqref{eq:P}, where, after the gauge transformation \eqref{eq:trafo}, $+(-)$ corresponds to the $S(A)$ state. Both $S$ and $A$ can then form a spin singlet- or triplet with the neighboring $d$-site.\\

For $t_{pd}\ll \Delta_{pd}<U_d$ and setting $t_{pp}^{(\prime)}=0$ for now, one can use second order perturbation theory in $t_{pd}/(U_{d}-\Delta_{pd})$, $t_{pd}/(U_{p}+\Delta_{pd})$ to determine the energies of the
singlet and triplet states: For $S$ states, the singlet ($s$) and triplet ($t$) energies are \cite{Zhang1988}
$$E_{Ss}=-8t_{pd}^2(\frac{1}{U_p+\Delta_{pd}}+\frac{1}{U_d-\Delta_{pd}})$$ 
and $E_{St}=E_{At}=0$; $A$ has energy~\cite{Zhang1988} $$E_{As}=-4t_{pd}^2\frac{1}{U_p+\Delta_{pd}}.$$
This leads to a gap $\Delta E=E_{As}-E_{Ss}$. We compare this prediction of $\Delta E$ to exact diagonalization (ED) results for two holes on a plaquette with one $d$-site and four surrounding $p$-sites. Fig.~\ref{fig:Gaps} (left) shows that the predicted perturbative gap (red line) agrees well with the spectrum from ED (second lowest blue line) for $t_{pp}=0$ and $t_{pd}\ll \Delta_{pd}<U_d$. For the perturbative treatment, we have assumed $t_{pp}=0$; however, the agreement for $t_{pp}\neq 0$ is still relatively good. For a more realistic parameter choice (Fig.~\ref{fig:Gaps}, right), the gap between the ground state and the first excited state from ED still agrees reasonably well with the perturbative prediction. Furthermore, we note that the gap from both ED and perturbation theory is large, on the order of $\geq1.5t_{pd}$.

\section{Band mixing fitting and Wannier state construction}

\begin{figure}[t]
    \centering
    \includegraphics[width=1\linewidth]{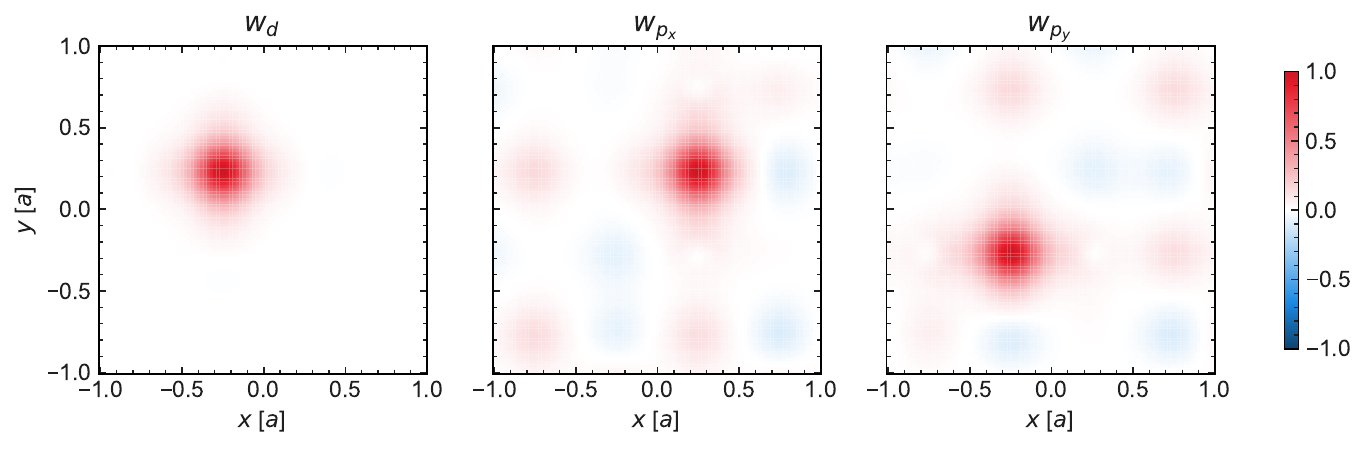}
    \caption{Wannier functions for $w_d$, $w_{p_x}$ and $w_{p_y}$ as shown in Fig~\ref{fig:1}\textbf{e}. $a$ is the unit cell spacing which is half of the wavelength and the grid is shifted by $a/4$ on each axes to match the unit cell coordinate axes. Here, we use $V=\SI{15}{E_r}$ and $\theta=\SI{0.025}{rad}$ as an example, and fitted $\Delta/t_{\rm pd} = 3.5$}
    \label{fig:wannier_function}
\end{figure}

We consider the Bloch eigenstates $\ket{\phi_{\vb q}^{(n)}}$ obtained by diagonalizing the single-particle Hamiltonian in the optical lattice (with potential given in Eq.~\ref{eq:Opticalpot}) on a uniform mesh of quasimomenta $\vb q$ spanning the first Brillouin zone (BZ). For each $\vb q$, the eigenproblem
\begin{equation}
  \tilde{\mathcal{H}}(\vb q)\ket{\phi_{\vb q}^{(n)}} = \varepsilon_{\vb q}^{(n)}\ket{\phi_{\vb q}^{(n)}}
\end{equation}
defines a set of band states labeled by $n$.

To construct Wannier states centered on orbital $\nu$ with $\nu = d,p_x,p_y$, we express the target state as a $\vb q$-space superposition of the Bloch eigenstates,
\begin{equation}
  \ket{w_{\nu}}=\frac{1}{\sqrt{N_s}}\sum_{\vb q\in \mathrm{BZ}}\sum_n s^{(n)}_{\vb q,\nu}\,e^{-i\theta^{(n)}_{\vb q}}\ket{\phi^{(n)}_{\vb q}}.
  \label{eq:wannier-state}
\end{equation}
where $N_s$ is the number of sampled $\vb q$ points. The mixing weights $s^{(n)}_{\vb q,\nu}$ specify how orbital $l$ is distributed over the band subspace, while the phases $\theta^{(n)}_{\vb q}$ fix a gauge that yields a localized real-space Wannier representation.

On the other hand, for the tight-binding describtion of the Emery model, one can also write down the Hamiltonian in momentum space $\tilde{\mathcal{H}}_{\rm TB}(\vb{q})$ with quasimomentum $\vb{q}=(q_x, q_y)$ as \cite{Tseng2025}

\begin{equation}
\begin{split}
        \mqty(-2t_{dd}\qty(\cos(q_x a) + \cos(q_y a)) & -2t_{pd}\cos\frac{q_xa}{2} & -2t_{pd}\cos\frac{q_ya}{2}\\
        -2t_{pd}\cos\frac{q_xa}{2} & \Delta_{pd}-2t_{pp^\prime}\qty(\cos(q_xa) + \cos(q_ya)) & -2t_{pp}\qty(\cos\frac{(q_x + q_y)a}{2}+\cos\frac{(q_x - q_y)a}{2}) \\
        -2t_{pd}\cos\frac{q_ya}{2} & -2t_{pp}\qty(\cos\frac{(q_x + q_y)a}{2}+\cos\frac{(q_x - q_y)a}{2}) & \Delta_{pd}-2t_{pp^\prime}\qty(\cos(q_xa) + \cos(q_ya))).
\end{split}
\end{equation}
If the optical-lattice bands are well captured by this tight-binding model, the hopping and onsite parameters can be optimized such that the eigenenergies of $\tilde{\mathcal{H}}_{\rm TB}(\vb q)$ match the numerically obtained $\varepsilon^{(n)}_{\vb q}$ for $n=1,2,3$.

Once the tight-binding parameters are determined, we also obtain the unitary transformation between the orbital basis $\qty(d,p_x,p_y)$ and the Bloch-band basis. This transformation directly provides the mixing weights $s^{(n)}_{\vb q,\nu}$. The remaining gauge freedom is fixed by choosing the phases such that the Bloch function at the site center $\vb r_\nu$ is real and positive:
\begin{equation}
    \theta_{\vb{q}}^{(n)} = \arg \phi_{\vb{q}}^{(n)}(\vb{r}_{\nu})
\end{equation}
With this choice, the Wannier functions follow from Eq.~\ref{eq:wannier-state}. As it shown in FIG.~\ref{fig:wannier_function}, we give an example of constructed Wannier functions with $V_0 = 15\, E_r$ and $\theta = \SI{0.025}{rad}$.

The on-site interaction $U_{\nu}$ for orbital $\nu\in\{d,p_x,p_y\}$ reads
\begin{equation}
U_\nu
= g \int d^2\mathbf r \, |w_\nu(\mathbf r)|^4\int \dd z\abs{w_\perp(z)}^4
\end{equation}
where $g=4\pi\hbar a_s/m$ is the interaction strength, with $a_s$ the $s$-wave scattering length (tunable via a Feshbach resonance \cite{chinFeshbachResonancesUltracold2010}), and $w_\perp(z)$ the Wannier function in the transverse direction. In this way, the overall scale of $U_\nu$ relative to the tunneling amplitudes or band gaps can be tuned straightforwardly. However, the \emph{relative} onsite interactions between the $p$ and $d$ orbitals are constrained by the orbital shapes through the overlap integral $\int d\mathbf r \, |w_\nu(\mathbf r)|^4$. The resulting ratios accessible in our proposed setup are shown in Fig.~\ref{fig:Experiment}. The accessible ratios $U_p/U_d\approx 0.6,\dots, 0.8$ are in good agreement with the values $U_p/U_d<1$ considered in the literature\cite{Hybertsen1989,Martin1996,Hanke2010,Cui2020,McMahan1988,White2015,Kent2008,MDopf1990,Jiang2023,Kowalski2021Oxygenholecontent,Tseng2025,SM}. Note that for small doping, there is only an average $p$-site density of $N_\mathrm{h}/2\ll 1$, and hence even for $U_p=0$ the probability of double-occupancies is very low. Consequently, the exact value of $U_p$ is not expected to change the physics of the system a lot.\\

\section{Experimentally accessible Hamiltonians}
In order to see which class of models we can access in principle, we perform a particle hole transformation $\hat{U}_{ph}$ with
\begin{align}
    \hat{U}_{ph}^\dagger \hat{d}_{\mathbf{i}\sigma}\hat{U}_{ph}=\hat{d}_{\mathbf{i}\sigma}^\dagger, \quad \quad \hat{U}_{ph}^\dagger \hat{p}_{\mathbf{i}\sigma}\hat{U}_{ph}=\hat{p}_{\mathbf{i}\sigma}^\dagger.
\end{align}
The transformed Hamiltonian becomes
\begin{align}
\hat{\mathcal{H}}^{ph} = & +t_{pd} \sum_{\langle \mathbf{i}\mathbf{j} \rangle \sigma} (\hat{d}^{\dagger}_{\mathbf{i}\sigma} \hat{p}_{\mathbf{j}\sigma} + \text{h.c.}) 
+ t_{pp} \sum_{\langle\langle \mathbf{j}\mathbf{j}^\prime \rangle\rangle \sigma} (\hat{p}^{\dagger}_{\mathbf{j}\sigma} \hat{p}_{\mathbf{j^\prime}\sigma} + \text{h.c.})
- \Delta_{pd} \sum_{\mathbf{j},\sigma} \hat{p}^\dagger_{\mathbf{j}\sigma} \hat{p}_{\mathbf{j}\sigma}
+ \sum_{\nu \in \{p,d\}} U_\nu \sum_\mathbf{i} \hat{n}^\nu_{\mathbf{i}\uparrow} \hat{n}^\nu_{\mathbf{i}\downarrow}.
\label{eq:3bandfinal}
\end{align}


\section{Numerical results}
\subsection{Convergence}
\begin{figure*}[t]   \includegraphics[width=0.95\textwidth]{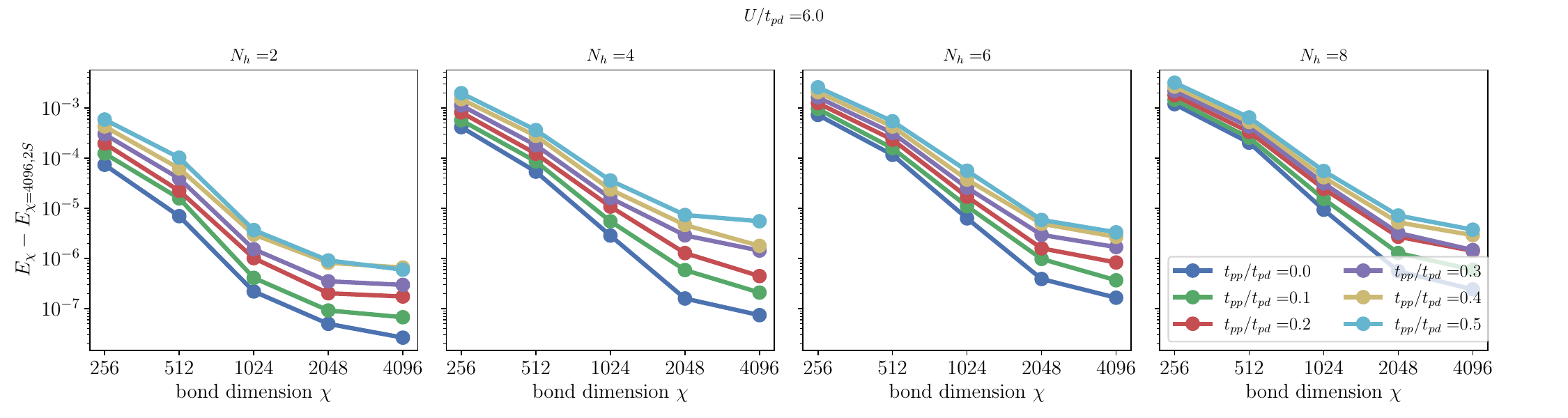}
\centering\includegraphics[width=0.95\textwidth]{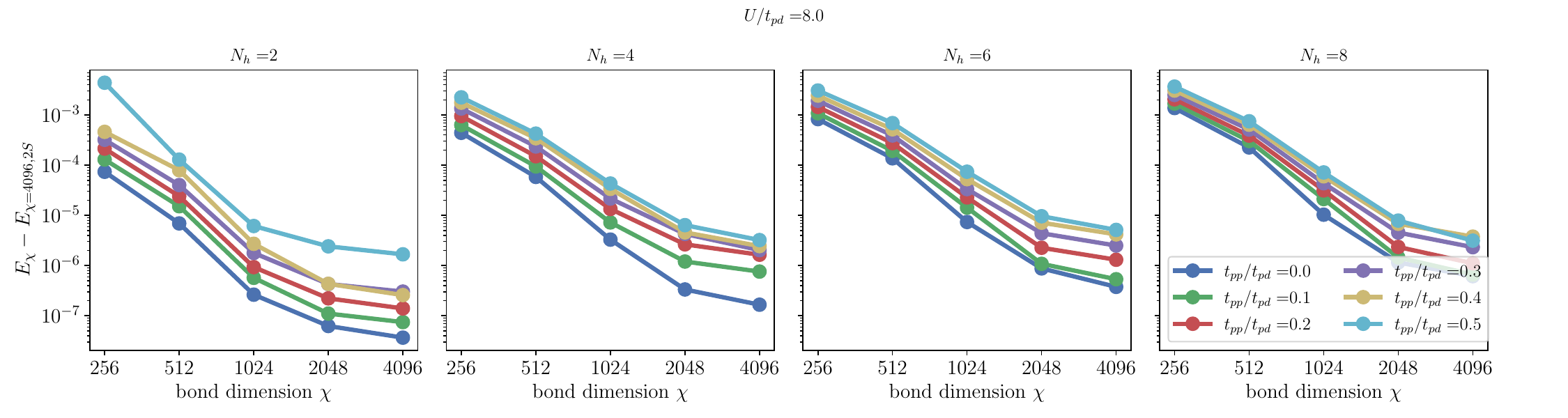}
    \caption{Energies $E_\chi$ for different bond dimensions $\chi$ for $12\times 2$ unit cells ($72$ sites) with periodic boundaries in the short direction. We consider different values of $U/t_{pd}$ (top and bottom panels), hole dopings $N_\mathrm{h}=2,4,6,8$ (left to right columns) and $t_{pp}/t_{pd}$ (different colors). The displayed energies are obtained using single-site DMRG and compared to the energy obtained with the highest bond dimension $\chi=4096$ and the two-site ($2S$) algorithm, $E_{\chi=4096,2S}$.}
    \label{fig:DMRGenergies}
\end{figure*}

All numerical calculations were performed using matrix product states implemented in the package SyTen \cite{syten1,syten2}, $U(1)$ symmetries corresponding to charge and $S_z$ spin conservation. For even (odd) particle numbers, we work in the $S_z = 0 \,(0.5)$ sector. We keep a maximum bond dimension of $\chi  = 4096$, optimizing with single-site density matrix renormalization group (DMRG) for the first four stages with bond dimensions $\chi=256,\dots ,4096$, and with two-site DMRG for the last stage with $\chi  = 4096$. The energies obtained after each stage for different Hamiltonian parameter choices and hole dopings are shown in Fig.~\ref{fig:DMRGenergies}. The maximum truncation error is $10^{-8}$.

\section{SU(2) invariance}
\begin{figure*}[t]   \includegraphics[width=0.8\textwidth]{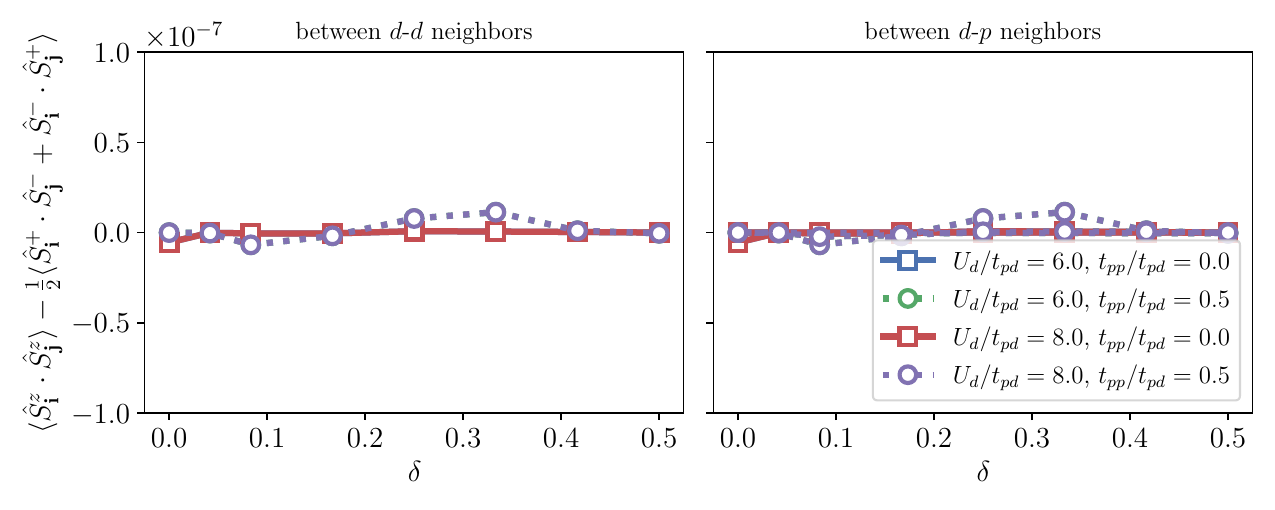}
    \caption{Difference of nearest nearest-neighbor spin correlations in $z$ direction and the $x,y$ plane, $\langle \hat{S}^z_\mathbf{i}\cdot \hat{S}^z_\mathbf{j} \rangle-\frac{1}{2}\langle \hat{S}^+_\mathbf{i}\cdot \hat{S}^-_\mathbf{j} +\hat{S}^-_\mathbf{i}\cdot \hat{S}^+_\mathbf{j} \rangle$ for a system of $L_x\times L_y=12\times 2$ sites and different dopings $\delta$.}
    \label{fig:SU2}
\end{figure*}

In the definition of $M_\mathrm{bond}$, Eq.~\eqref{eq:Mbond}, we assume that the ground state is SU(2) invariant. Fig.~\ref{fig:SU2} indicates that this SU(2) invariance is indeed present in the numerically obtained ground states by considering the difference between nearest-neighbor spin correlations in $z$ direction and the $x,y$ plane, $\langle \hat{S}^z_\mathbf{i}\cdot \hat{S}^z_\mathbf{j} \rangle-\frac{1}{2}\langle \hat{S}^+_\mathbf{i}\cdot \hat{S}^-_\mathbf{j} +\hat{S}^-_\mathbf{i}\cdot \hat{S}^+_\mathbf{j} \rangle$.

\end{document}